\documentclass[5p,times]{elsarticle}

\usepackage{lineno,hyperref}
\usepackage[utf8]{inputenc}
\usepackage{graphicx}
\usepackage{graphics}
\usepackage{float}
\usepackage{siunitx}
\usepackage{blindtext}
\usepackage[nice]{nicefrac}
\usepackage[dvipsnames]{xcolor}
\usepackage{amsmath}
\usepackage{lscape}

\usepackage{draftwatermark}
\SetWatermarkText{}
\SetWatermarkScale{4}
\SetWatermarkColor[gray]{0.87}

\newcommand*{\ditto}{\texttt{"}}

\graphicspath{{./figures/}}

\modulolinenumbers[5]

\journal{arXiv}

\begin{document}

\begin{frontmatter}

\title{On the risk of infection by infectious aerosols in large indoor spaces}

\author[1]{Bardia Hejazi}

\author[1]{Oliver Schlenczek}

\author[1,2]{Birte Thiede}

\author[1]{Gholamhossein Bagheri}

\author[1,2,3]{Eberhard Bodenschatz\corref{mycorrespondingauthor}}
\cortext[mycorrespondingauthor]{Corresponding authors}
\ead{eberhard.bodenschatz@ds.mpg.de}

\address[1]{Laboratory for Fluid Physics, Pattern Formation and Biocomplexity, Max Planck Institute for Dynamics and Self-Organization (MPIDS), Göttingen, NI 37077, Germany}
\address[2]{Institute for Dynamics of Complex Systems, University of Göttingen, Göttingen, NI 37077, Germany}
\address[3]{Laboratory of Atomic and Solid State Physics and Sibley School of Mechanical and Aerospace Engineering, Cornell University, Ithaca, NY 14853, USA}
\address[4]{Sibley School of Mechanical and Aerospace Engineering, Cornell University, Ithaca, NY 14853,USA}

\begin{abstract}

Airborne diseases can be transmitted by infectious aerosols in the near field, i.e., in close proximity, or in the far field, i.e., by infectious aerosols that are well mixed within the indoor air. Is it possible to say which mode of disease transmission is predominant in large indoor spaces? We addressed this question by measuring the transport of aerosols equivalent to the size of human respiratory particles in two large hardware stores ($V>\SI{10000}{\cubic m}$). We found that aerosol concentrations in both stores decreased rapidly and almost independently of aerosol size, despite the different ventilation systems. A persistent and directional airflow on the order of a few cm/s was observed in both stores. Consequently, aerosol dynamics in such open settings can be expected to be dominated by turbulent dispersion and sweeping, and the accumulation of infectious aerosols in the indoor air is unlikely to contribute significantly to the risk of infection as long as the occupancy of the store is not too high.  Under these conditions, well-fitting face masks are an excellent means of preventing disease transmission by human aerosols.

\end{abstract}


\end{frontmatter}


\section{Introduction}

It is important to keep the indoor air we breathe free of pollutants and other harmful substances, as we spend most of our time indoors~\citep{Jenkins1992, Klepeis2001}. Airborne infectious diseases such as Severe Acute Respiratory Syndrome (SARS), avian flu and swine flu (H1N1), and more recently the coronavirus disease (COVID-19) are transmitted through near-field and far-field airborne contact with an infectious ~\citep{Zhang2020,WHO2020, Pohlker2021}; with the predominant route of transmission for SARS-CoV-2 being via airborne transport of respiratory aerosols released from the respiratory tract of an infectious person~\citep{Leung2021}. (In this study, we use the terms \emph{particle(s)} and \emph{aerosol(s)} interchangeably to refer to \textless\SI{100}{\micro\meter} particulate matter suspended in air, regardless of composition.) The size of human aerosol emissions varies greatly in magnitude and span length scales of several decades~\citep{Bagheri2021D}.
While moist exhaled aerosols larger than \SI{50}{\micro\meter} (approx. \SI{10}{\micro\meter} fully dry~\citep{Bagheri2021D})  settle within seconds by gravitational deposition, smaller droplets dry while airborne and shrink in size,  and due to their small remaining mass can stay airborne for extended periods of time~\citep{Pohlker2021}. Such aerosols may contain single or multiple copies of pathogens when exhaled by an infectious individual, and when inhaled by a susceptible person, there is a risk of infection determined by the absorbed dose of pathogens~\citep{Nordsiek2021}.

\begin{table*}[!htb]
\fontsize{8}{11}\selectfont
\centering

\begin{tabular}{c c c c c c}
 \multicolumn{6}{c}{} \\
\hline

author &
room volume ($\text{m}^{3}$) &
ACH ($\text{hr}^{-1}$) &
$\tau$ (min)&
measurement duration (min)&
description

\\\hline

\citet{Ishizu1980} & 71 & NA & 7.66 & 21 &  cigarette smoke, ventilated room\\
\ditto & 268 & NA & 6.83 & 21 & \ditto \\
\ditto & 82 & 45.3 & 7.9 & 15 & outside air intake rate \\
\ditto & 82 & 8.7 & 8.68 & 22 & \ditto \\
 
\citet{Leaderer1984} & 34 & 2.3 & 12.9 & NA &  cigarette smoke in room\\

\citet{Mage1996} & 548 & NA & 2.6 & 32 & CO concentration, cigarette in Tavern \\

\citet{Miller2001} & 36 & 0.03-1.7 & 23-40 & 150-240 & cigarette smoke, 2 compartments \\

\citet{Ott2003house} & 34 & 4 & 44 & 225 & CO concentration, cigar smoke,  2 compartments \\

\citet{Ott2008cars} & 3-5 & 3-56.4 & 0.3-7.8 & NA &  $<2.5 \mu \text{m}$ particles, cigarette in car \\

\citet{Qian2012} & 242 & NA & 43 & 70 &  $<10 \mu \text{m}$ particles in 2 story house \\

\citet{Stephens2013} & 45 & NA & 23.6 & 59 & $10-100 \text{nm}$ particles, 3D printer emissions,\\  & & & & & conditioned air \\

\citet{Poon2016} & 11.9 & 40.6 & 11.8 & 32 & $<100 \text{nm}$ particles, cooking in source room, \\
 & & & & & 2 compartments\\

\hline

\end{tabular}

\caption{Examples of experimental studies on indoor pollutants and their decay.}
\label{table:lit}
\end{table*}

Far-field transmission in indoor air is described by the well-mixed room (or space) model and its extensions. These models assume that the indoor air is well mixed, in other words, susceptible individuals who can be infected are far enough away from the infectious so that the infectious aerosols are homogeneously  mixed into the considered volume  by the time they are inhaled by the susceptible \citep{Miller_2020, Nazaroff_1998, Fennelly_1998,Lai_2000, riley2002indoor,Nicas_2005, He_2005, Miller_2020,BUONANNO_2020}. The literature on well-mixed models is substantial, i.e., the spaces considered include aircraft cabins~\citep{Wan2009}, buses~\citep{Zhu2012}, hospital wards~\citep{Qian2010,Ren2021}, and residential buildings~\citep{He2005,Gao2009}, or recent case study scenarios of different indoor settings~\citep{Kurnitski2021}. Experiments have investigated the transport dynamics of pollutants in differently sized  indoor spaces~\citep{Ishizu1980, Hussein2006,Ott2008cars} and how human presence and activity can affect these dynamics~\citep{Qian2012,Licina2014,Cheng2021}. Studies have also looked at how aerosols can transfer from room to room~\citep{Lai2008,Ott2003house,Poon2016}. Best strategies for ventilation of indoor spaces to minimize disease transmission have been devised for a long time, recent examples are  for   schools~\citep{Curtius2021}, hospitals and clinics~\citep{Mousavi2021,Ren2021b}, and homes~\citep{Alavy2020}.  The well-mixed model fails when the volume of the indoor space is large or when the infectious and the susceptible are  close to each other. Extensions for such situations are by subdividing the indoor environment into near- and far-field ~\citep{Nicas_2008, Cherrie_2011,Arnold_2017} or performing situation-specific approximate numerical simulations with turbulent mixing models ~\citep{Gao2008,Zhao2009,Sajjadi2016}. Aerosol concentration decay in well-mixed rooms can be well described with an exponential decay. Typical measured exponential time constants under various air exchange conditions (room volume, air change hours - or ACH, and measurement duration) are given for later comparisons with the results reported here in Table~\ref{table:lit}.

For near-field disease transmission the temporal dynamics of aerosol dispersion is highly dependent on the activity-duration of the source and the dispersion mechanism. \citet{Mage1996} introduced three time-scales for the aerosol concentration evolution, which were further investigated by \citet{Klepeis1999} and \citet{Licina2017}. Namely, the time when the source is active $t_{\alpha}$, the time when the source is no longer active and the space is not well-mixed $t_{\beta}$ , and the time from when the space is well-mixed to when the aerosol concentration reaches that of the background $t_{\gamma}$. For  $ t_{\alpha}+t_{\beta} \ll t_{\gamma} $ the well-mixed room model can be considered  adequate  for exposure risk calculations~\citep{Mage1996}. \citet{Bagheri2021} investigated the upper bound on the risk of disease transmission for one-to-one near-field exposure of human exhale in different scenarios of social distancing for unmasked and masked individuals and found that wearing a well-filtering face mask is the safest option and greatly reduces the infection risk. Recently, \citet{Li2021} have investigated the effect of ventilation on the transmission by respiratory aerosols in well-mixed spaces by considering contaminant concentration in the near-field exhalation cone of an individual. It is also important to note that  exhaled human aerosols can remain airborne for longer periods of time than previously thought~\citep{Wang2021}.

Large room volumes have found little attention in the literature so far as shown in Table \ref{table:lit}. \citet{Mage1996} investigated  a tavern with $V\sim \SI{550}{\cubic\meter}$ where they report relatively short measured decay times of $\sim 3$min with no information on the ventilation. \citet{Blocken2021}  examined  how different aerosol removal mechanisms effect aerosol concentrations of particles emitted by persons performing physical exercises in a gym with $V\sim \SI{900}{\cubic\meter}$. They found that a combination of ventilation and air filtering units work best in reducing aerosol concentration. They reported total concentrations at specific time intervals, but did not report  decay times for aerosol concentration.  Comparing the results by \citet{Ishizu1980} and \citet{Qian2009} shows that knowing the room volume alone is not sufficient to estimate the aerosol concentration decay time. These investigations  studied two similar sized rooms, $\SI{268}{\cubic\meter}$  and $\SI{242}{\cubic\meter}$  and found quite  different  decay times of $\sim 7$ min versus  $\sim 40$ min, respectively.  \citet{Ishizu1980} also compared the same room with different ACH values of 45 and 9 and found decay times of $\sim 8$ min for both scenarios. This was attributed to non-ideal mixing of the clean air supply with the existing room air. It is remarkable that similarly puzzling results were found when comparing \citet{Leaderer1984} with \citet{Ott2003house} where similar rooms have quite different decay rates. In \citet{Ott2003house} for the very small $\SI{34}{\cubic\meter}$ room studied, room compartmentalization and transport from room to room increased local decay times in spite of the ACH being almost doubled. These examples (listed in Table \ref{table:lit}) show that predicting decay times in an indoor environment based solely on reported values of ACH and room volume is a non-trivial task and may lead to significant uncertainties.

The importance of the room airflow on aerosol transport dynamics was recently pointed out by~\citet{Bhagat2020}. Detailing airflow in rooms is challenging, if not impossible, due to the complexity introduced by boundary conditions, placement of air-inlets and air-outlets, opening and closing of doors, and other time dependent factors such as the movement of people. \citet{Bhagat2020} suggest that the room flow becomes a key component in aerosol transmission of COVID-19 for particle sizes that have gravitational settling speeds lower than the typical room flow velocities.  In the following, our data shows that room flow velocities in the two measured exemplary hardware stores are fast, robust to spatio-temporal fluctuations and that settling of particles is negligible. In other words aerosol concentration decay rates are short, as shown later.   Thus the infection risk will be dominated by near-field  transmission for which we can use our earlier results to calculate an upper bound for the risk of infection \citep{Bagheri2021D}.  From this upper bound, we conclude that presence in these large indoor spaces poses a low risk of infection for current variants of SARS-CoV-2 as long as well-fitted filtering face masks are worn and room occupancy is limited.

\section{Measurements of aerosol decay and transport in two large hardware stores}

\begin{figure*}[htbp]
	\centering
	\includegraphics[width=1\textwidth]{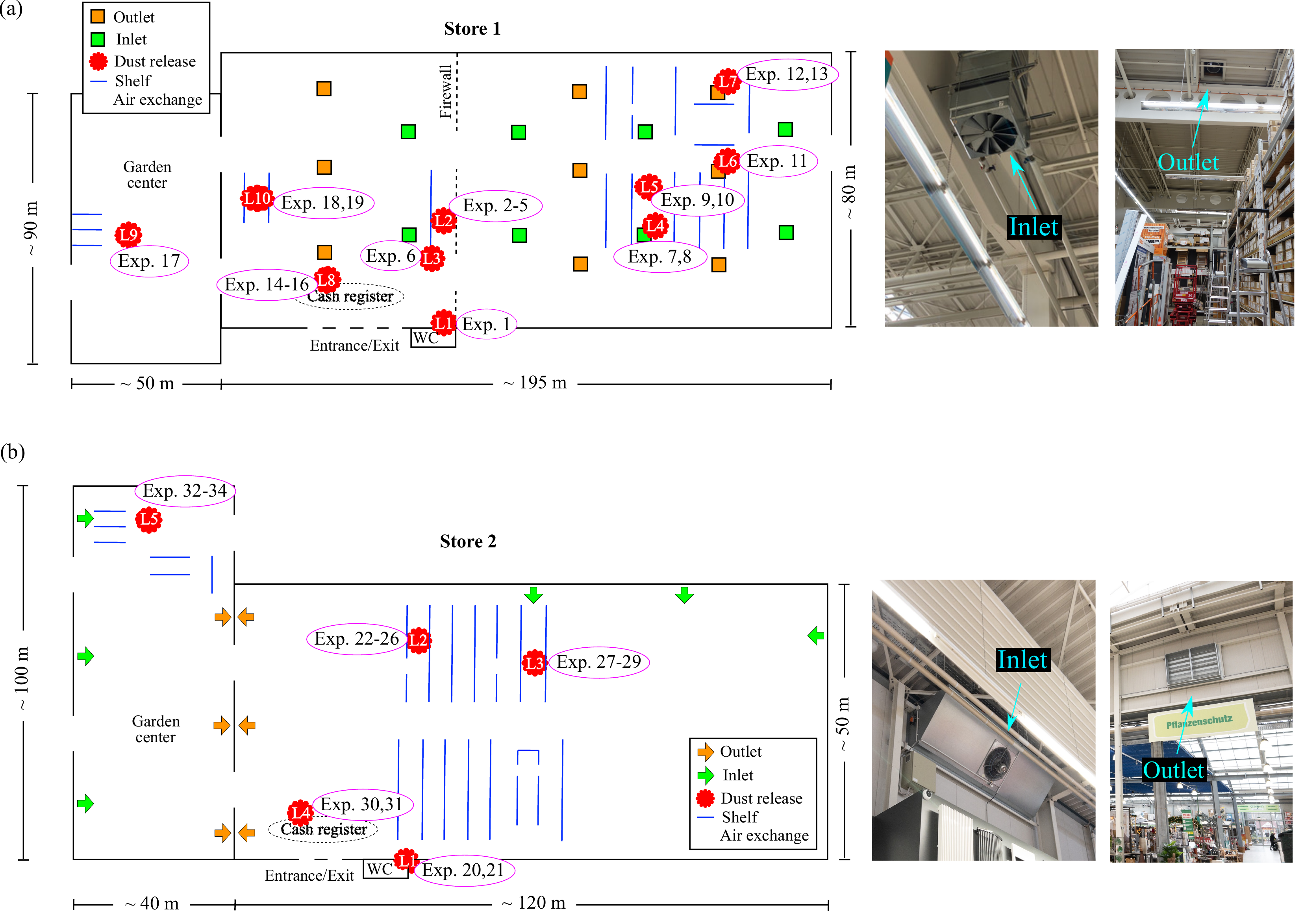}
	\caption{Schematic top view of the store layout with ventilation systems, as well as entrances and exits that provide additional air exchange with the outside air. The locations of test aerosol release and aerosol measurements are also marked with location numbers and number of experiment performed in a particular location. To reduce complexity, only the shelving around the experiment locations are shown.}
	\label{locations}
\end{figure*}

We performed measurements at two different hardware stores in Germany during the COVID-19 pandemic and while the stores were open to pre-registered customers with limits on store capacity. All individuals present in the store were wearing filtering face masks and practicing social distancing in compliance with local hygiene rules and regulations. As a result the stores had low occupancy and people were not present during measurements unless measurements were performed specifically to study the effects of human presence and movement. Both stores had the layout typical to large chain hardware stores with high ceilings (\SI{\sim10}{\meter}) and large interiors but different ventilation layout as shown in Fig.~\ref{locations}. 

\textbf{Store 1} had a total volume of approximately $\sim$\SI{200000}{\cubic\meter}, while the semi-detached garden center had a volume of $\sim$\SI{45000}{\cubic\meter}. The garden center was an enclosed greenhouse and had less shelving and more open space. 
In Store 1 (as shown in Fig.~\ref{locations}(a)) the ventilation units (inlets and outlets) are spread over the entire retail area at regular intervals on the ceiling, so that in a given section of the store, a row of inlet units followed a row of outlet units. Inlets supply air into the indoor space and outlets extract air from the indoor space. 
The main retail area, excluding the garden center, had a maximum air exchange rate of 57800 cubic meters per hour provided by the air handling units, which results in a nominal ACH of \SI{0.4}{\per \hour}. The store's air exchange rate was reported to us by building services.
The store had 300 total parking spaces for cars and with an average of 2 people per vehicle~\citep{GFK2020}, the estimated maximum occupancy is $\sim$\SI{32}{\square \meter}/person.

\textbf{Store 2} had a volume of $\sim$\SI{60000}{\cubic\meter} in the main retail area and $\sim$\SI{40000}{\cubic\meter} in the garden center. For Store 2, the ventilation units were installed on the walls of the store wile the outlets were on the wall separating the main retail area and the garden center (see Fig.~\ref{locations}(b)).
The inlets of Store 2 were close to the ceiling and in the main retail area they blew air into the store at a downward angle of $~45^{\circ}$, as seen in Fig.~\ref{locations}(b), while in the garden center they blew air perpendicularly away from the wall (Fig.~12 of the Supplementary Material). 
The air exchange rate was reported to us by building technicians. 
The main retail area had 3 fans that generated an air exchange rate of \SI{12000}{\cubic\meter \per \hour} each, resulting in a nominal ACH of \SI{0.6}{\per \hour} for the whole retail area. The garden center also had 3 fans with an air exchange rate of \SI{7000}{\cubic\meter \per \hour} for each fan, which in total generate a nominal ACH of \SI{0.5}{\per \hour}.
As before, the maximum occupancy of the store is estimated to be $\sim$\SI{27}{\square \meter}/person, from the total 180 available parking spaces and an average number of two people per vehicle.

The values for occupancy based on parking spaces are the extreme maximum and actual store occupancy as reported by store officials is less, where on normal operating days the occupancy is $\sim$\SI{100}{\square \meter}/person, and during promotional periods the occupancy is $\sim$\SI{60}{\square \meter}/person. Furthermore, according to internal reports, more than $95\%$ of customers stay in these stores for less than one hour, with more than 50\% staying for less than half an hour~\citep{GFK2020}. The customer toilets of both stores were small spaces with a volume of approximately \SI{15}{\meter^{3}}. By German law the  ACH of the customer toilets should be at least \SI{5}{\per \hour}~\citep{toilet_law}. 

\begin{figure*}[htbp]
	\centering
	\includegraphics[width=1\textwidth]{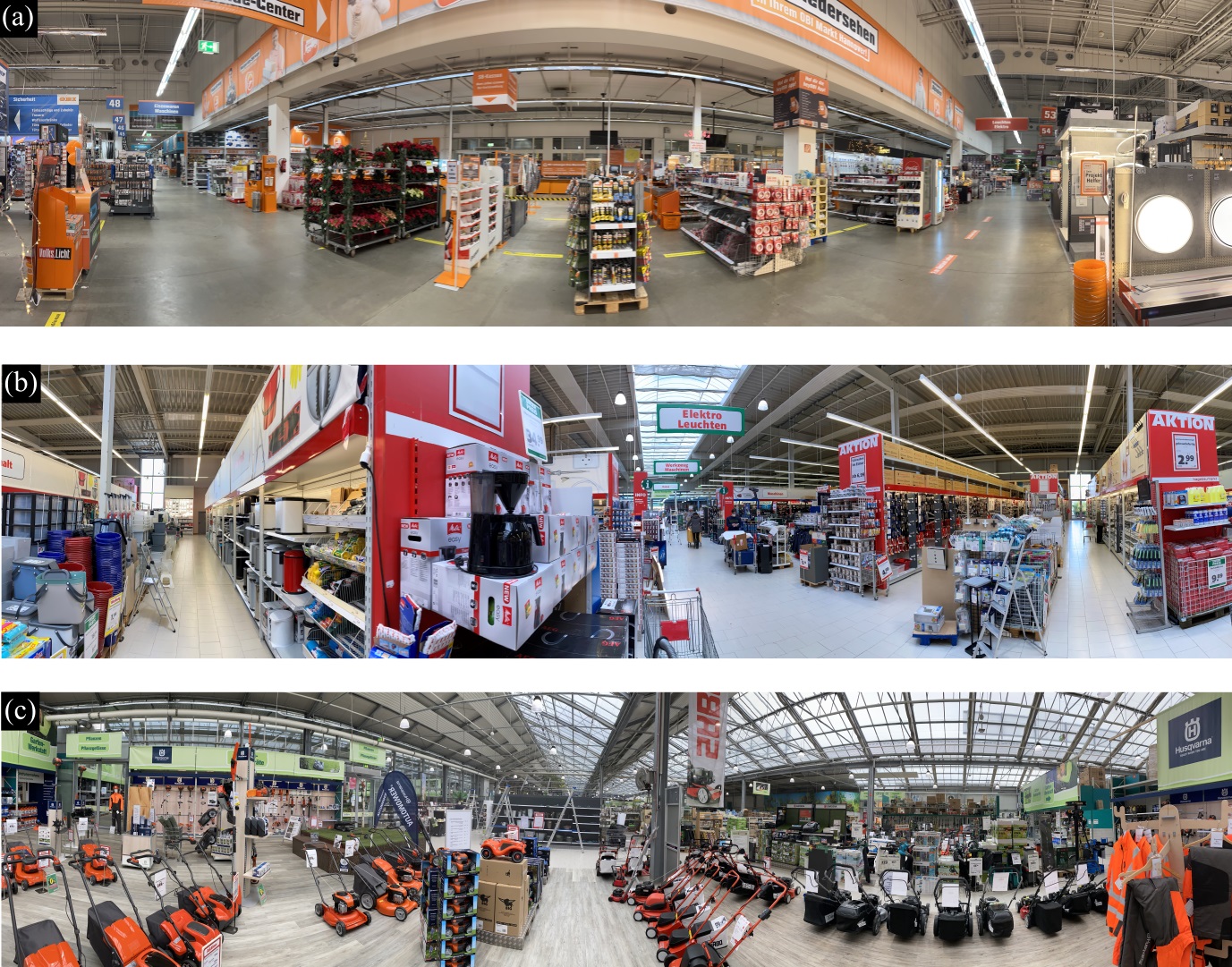}
	\caption{Panoramic photographs of the stores. (a) The main retail area of Store 1 and (b) Store 2, showing the large interior space with high ceilings. (c) The garden center of Store 2 which is a greenhouse with less shelving and more open space as compared to the main retail area. The garden center of both stores were very similar.}
	\label{photos}
\end{figure*}

Figure~\ref{locations} also shows the locations where the test aerosol was released and the aerosol measurements took place labeled by location number and additionally showing the experiment number that was performed in that particular location. The measurement locations were carefully selected, e.g. the screw and nail departments, where there can be a high number of customer visits and customers dwell longer.  The location of the shelves near the test sites is also shown in Fig. ~\ref{locations}, as shelves can influence aerosol transport. The Supplementary Material contains a detailed report of each experiment, showing the locations of the aerosol release and measurement devices, including the layout of the surrounding shelves and supporting photographs.
Fig.~\ref{photos} shows panoramic photographs of the main retail area of both stores and the garden center of Store 2. The photographs display the large interiors and high ceilings of the stores. The garden centers of both stores were very similar with less shelving and more open space as compared to the main retail area.

\begin{figure}[htbp]
	\centering
	\includegraphics[width=0.5\textwidth]{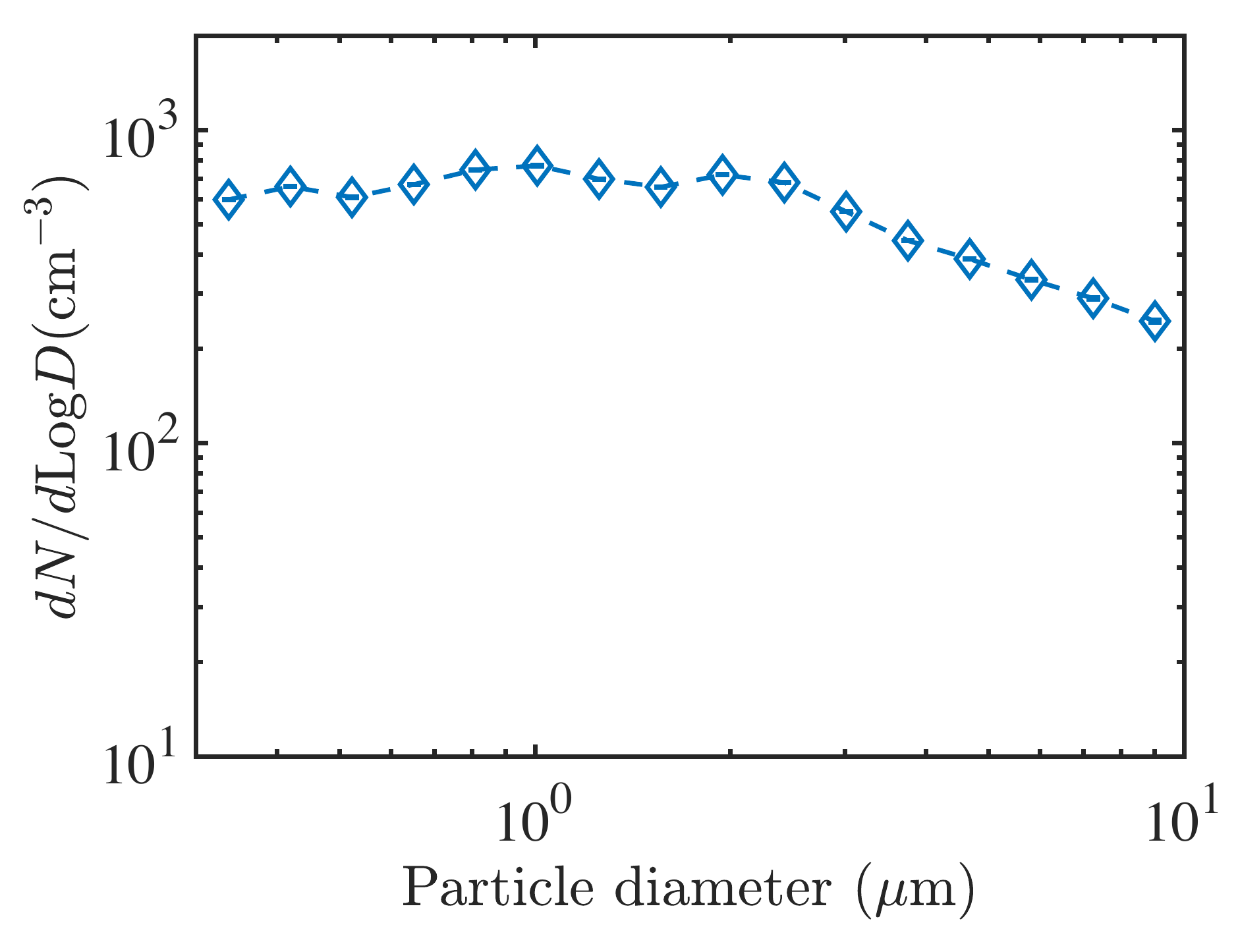}
	\caption{Size distribution of the DMT test dust measured immediately after the release with an spectrometer.
}
	\label{DMT}
\end{figure}

The aerosol measurements were conducted with four Optical Particle Sizer spectrometers TSI 3330, all of which were fully calibrated, and had valid calibration certificates.
The individual spectrometers were labeled as position 1, 2, 3 and 4 in each of the experiments. At the beginning of data collection all spectrometers were synchronized with each other in time. In each location they were placed on ladders or shelves at approximately the typical breathing height of an adult human (about $1.5$-\SI{2.0}{\meter} above the ground). The background concentration in the store was measured at the beginning and end of each experiment. The spectrometers can measure particle sizes in the range of $0.3$-\SI{10}{\micro \meter}. The spectrometers allow for a maximum of 16 custom selected bins for size distribution analysis. In this investigation we used 16 bins that were evenly spaced on the logarithmic scale. The sampling time was set to 1 second, which yields a sample volume of \SI{16.7}{\centi \meter^3} per scan.  The measured range does not cover the smaller aerosols of human origin,  however, this is not an issue since aerosols with size of approximately \SI{0.3}{\micro\meter} have the longest dwell time in ambient air, longer than smaller or larger particles.  We used dolomite test dust from DMT GmbH \& Co KG with diameters less than \SI{20}{\micro \meter}. The aerosol size distribution is almost flat over the measured size range and is shown in Fig.~\ref{DMT}. The test-dust is harmless to humans at the concentrations used, and furthermore all persons exposed to the dust wore FFP2 face masks.

The test dust was released at a height of about 1.0-\SI{1.5}{\meter} above the ground by first covering both sides of a \SI{30}{\centi\meter} x \SI{30}{\centi\meter} microfiber cloth with dolomite dust and then shaking it in an empty bucket with a volume of about \SI{30}{\liter} for a total duration of \textless\SI{10}{\second}. An example of the test-dust release can be seen in Fig.~14 of the Supplementary Material, we have also provided a video of the release. The utility of this simple method is also reflected by the fact that repeated experiments gave equal results on the decay times showing that the variability of the procedure had no significant impact on the results. After the local dust release, we monitored the time evolution of the concentrations at the four spectrometer locations. Once the measured aerosol concentrations decreased to the background level at all spectrometers, the experiment was terminated.

Two questions about the quality of the test dust measurements come immediately to mind:\newline
\textit{Could the test-dust itself influence the flow of the air?}\newline
The volume fraction of particles is defined as $\Phi_{p}=N V_{p}/V$, where $N$ is the total number of particles, $V_{p}$ is the volume of a single particle, and $V$ is the volume occupied by particles and fluid. To calculate an upper limit for $\Phi_{p}$, even if we assume that all particles are \SI{5}{\micro \meter}, we find that from the concentration data obtained by the particle counters, in each release 
the particle volume fraction was $\Phi_{p}<10^{-7}$, which is well within the dilute one-way coupling regime~\citep{Elghobashi1994}. Thus the flow influences the motion of particles, but the particles do not influence the flow or each other.

\textit{Does the procedure of release of the test dust influence the measured aerosol concentration decay?}\newline
The decay time for the turbulent kinetic energy generated by the cloth motion is approximated as $t=L/U$, where $L$ is a characteristic length scale and in our experiments this is the opening of the bucket ($L=\SI{0.3}{\meter}$), and $U$ is the mean flow velocity.
From videos of the dust release (see video provided in the Supplementary Material), we find that in our experiments the mean flow velocity created by the motion of the cloth is approximately $U=\SI{1}{\meter\per\second}$. Thus the kinetic energy of turbulence created by the dust generation decays in $t\sim\mathcal{O}(\SI{1}{\second})$, which is much shorter than the measured mean aerosol cloud decay times of about 1-2 minutes. From this we conclude that the turbulence generated by the cloth is suitable for creating a test dust cloud and has no influence on the transport and decay time measurements. All in all, this shows that the release of test dust by the bucket method provides reliable and reproducible results when the turbulent large-scale dispersion of aerosols is considered (see the Supplementary Material section 2 for additional details on the reproducibility of particle concentrations with aerosol release). 

Additionally, to determine whether thermally induced convective currents could impact aerosol transport, the temperature distribution at the measurement locations in one of the stores (Store 1) were measured with a thermal imaging camera (VarioCam head from Jenoptik Laser Optik Systeme GmbH, equipped with an IR 1.0/25 LW lens).

\begin{figure*}[htbp]
	\centering
	\includegraphics[width=0.9\textwidth]{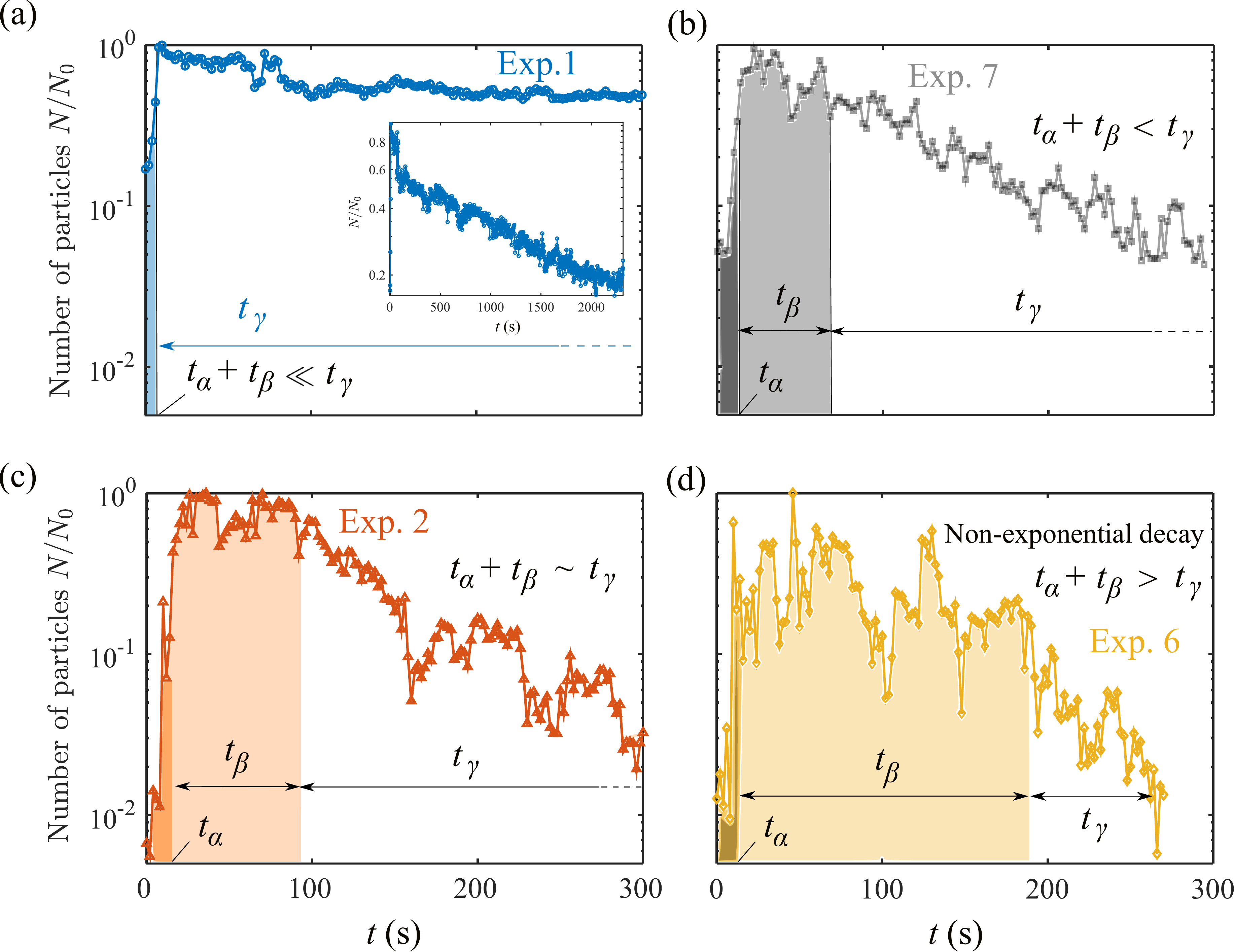}
	\caption{Normalized aerosol concentration for 0.90-\SI{1.12}{\micro\meter} sized particles as measured by one spectrometer for the first \SI{300}{\second} after aerosol release in Store 1. (a) Experiment 1 with an exponential decay where $ t_{\alpha}+t_{\beta} \ll t_{\gamma} $, displaying well-mixed room dynamics. The inset shows the time it takes for the concentration to decay and reach the initial background concentration. (b) Experiment 7 and (c) Experiment 2 where $t_{\alpha}+t_{\beta}$ is comparable to $t_{\gamma}$ yet an exponential can be fitted to the decay to calculate an effective $\tau$. (d) Experiment 6 where no good estimates for $\tau$ are found but aerosol dwell time $t_{90\%}$ can still be measured. }
	\label{exp_examples}
\end{figure*}


\textbf{Aerosol decay times}

\citet{Mage1996} proposed to analyze mixing/dispersion, and decay of aerosol concentration in similar experiments to ours  with  the times $t_{\alpha}$, $t_{\beta}$ and $t_{\gamma}$.  In Fig.~\ref{exp_examples} we present representative examples from our measurements and show the times $t_{\alpha}$, $t_{\beta}$ and $t_{\gamma}$ in order to define later the aerosol decay times used in the analysis of our measurements.  The data for the spectrometer bin  0.90-\SI{1.12}{\micro\meter} are shown as that bin had the best  signal-to-noise ratio. The aerosol rise time  $t_{\alpha}$ is $\sim$\SI{10}{\second}, $t_{\beta}$ is the time where the concentration approximately stays constant after release and before the decay starts, and $t_{\gamma}$ is the time from when the concentration starts to decay until it reaches that of the background. 
Figure~\ref{exp_examples}(a) shows the concentration decay of Experiment 1, which was performed in the customer toilet of \textbf{Store 1} with  one spectrometer. It can be seen that the decay timescale is much larger than the other timescales involved ($t_{\alpha}+t_{\beta}\ll t_{\gamma}$) where $t_{\gamma}$ extends well beyond \SI{300}{\second}, as shown in the inset. This shows that this customer toilet can be considered a well-mixed space, and as a result, the concentration decay can be fitted well with an exponential (with $0.76<r^{2}<0.97$ for all channels of the spectrometer).   The example shown in  figure~\ref{exp_examples}(b) is  experiment 7 in the retail space of \textbf{Store 1} (spectrometer Pos. 2 in Fig.~2 of the Supplementary Material), which was conducted under a ventilation inlet. Here $t_{\alpha}+t_{\beta} < t_{\gamma}$ and  $ t_{\gamma}$ can be fitted with an exponential ($0.54 <r^{2}<0.83$).  Experiment 2, shown in Fig.~\ref{exp_examples}(c) (spectrometer Pos. 1 in Fig.~1 of the Supplementary Material), was conducted in a relatively narrow  aisle in \textbf{Store 1}.  Fluctuations of concentrations occur during  mixing/dispersion $t_{\beta}$ and also during the decay $t_{\gamma}$.  Nonetheless, if we do not consider the highly fluctuating parts of the $t_{\gamma}$ an exponential fit with $0.90<r^{2}<0.94$ can be achieved. Figure~\ref{exp_examples}(d) shows the aerosol concentration measured for Experiment 6 of \textbf{Store 1} (spectrometer Pos. 4 in Fig.~1 of the Supplementary Material), conducted in a relatively wide and open area near the entrance/exit, which shows strongly fluctuating concentration values. For this experiment the $t_{\beta}$ and $t_{\gamma}$ regions have an exponential fit with $0.25<r^{2}<0.75$ (median $r^{2}\sim0.45$). As these examples show, for aerosol concentration decay the most important time scale is $t_{\gamma}$. In the analysis of our data we used two ways to measure $t_{\gamma}$, which we term the \emph{exponential aerosol decay time $\tau$} and  the \emph{aerosol dwell time $t_{90\%}$}.  These two parameters are correlated and determine how fast the aerosols are removed from the indoor air. 

We calculate the \textbf{\emph{exponential aerosol decay time $\tau$}} by fitting an exponential decay to the spectrometer measurements, where we report $\tau$ for experiments with $r^2 \geq 0.7$ (median $r^2\sim0.8$) for at least $80\%$ of the spectrometer bins. Overall, 42\% of all measurements (54 of 127) met these stringent requirements. Within the part of the time series where the exponential fit was applied, the change in concentration was at least a factor of 1.8 (median: factor 7.5); the smallest change in concentration was found in the experiments at the cash register of Store 2, the toilets in both stores, and the garden center of Store 1. The length of the data-fit for 50 out of the 54 measurements was at least one unit of $\tau$.

Exponential fits could not be applied to highly fluctuating decays, such as the experiment shown in Figure~\ref{exp_examples}(d), and therefore we define as a second parameter the  \textbf{\emph{aerosol dwell time $t_{90\%}$}} which we were able to find  for all measurements. $t_{90\%}$ is  the time it takes for the cumulative normalised concentration to go from 0.05 (5th percentile) to 0.95 (95th percentile). Using other thresholds is possible, but this choice worked well for quantifying all data measured.

In addition, we calculate the \textbf{\emph{aerosol advection velocity $v$}}.
For this we use the 0.05 value of the normalized cumulative aerosol concentration to calculate the aerosol time of arrival at a spectrometer. By knowing the distance between the spectrometers and aerosol release and dividing this value by the aerosol time of arrival, we calculate  $v$ (details of all spectrometer and aerosol release locations of all experiments are in the Supplementary Material).

\section{Results}

In total, we conducted 31 experiments with four spectrometers each, and three experiments with only one spectrometer. This yields a total of 127 data sets. It should be noted that in some experiments, the dust did not reach all spectrometer locations due to large-scale flows in the stores. And because of that, we needed to define a concentration threshold for those data sets that got a significant amount of dust. We calculated the maximum total concentration for each experiment and compared the total concentration at each of the other spectrometer locations with this value. If the total concentration at those locations was at least 5 \% of the maximum total concentration of the experiment, the respective data sets were considered significant and used for further analysis of decay dynamics. This was the case for 104 out of the 127 data sets. The experiment where the maximum aerosol concentration compared to the background aerosol concentration was the lowest, was still above the background by a factor of 27, so the maximum total concentration used to define the threshold was never close to or below the background concentration.

\begin{figure*}[htbp]
	\centering
	\includegraphics[width=0.9\textwidth]{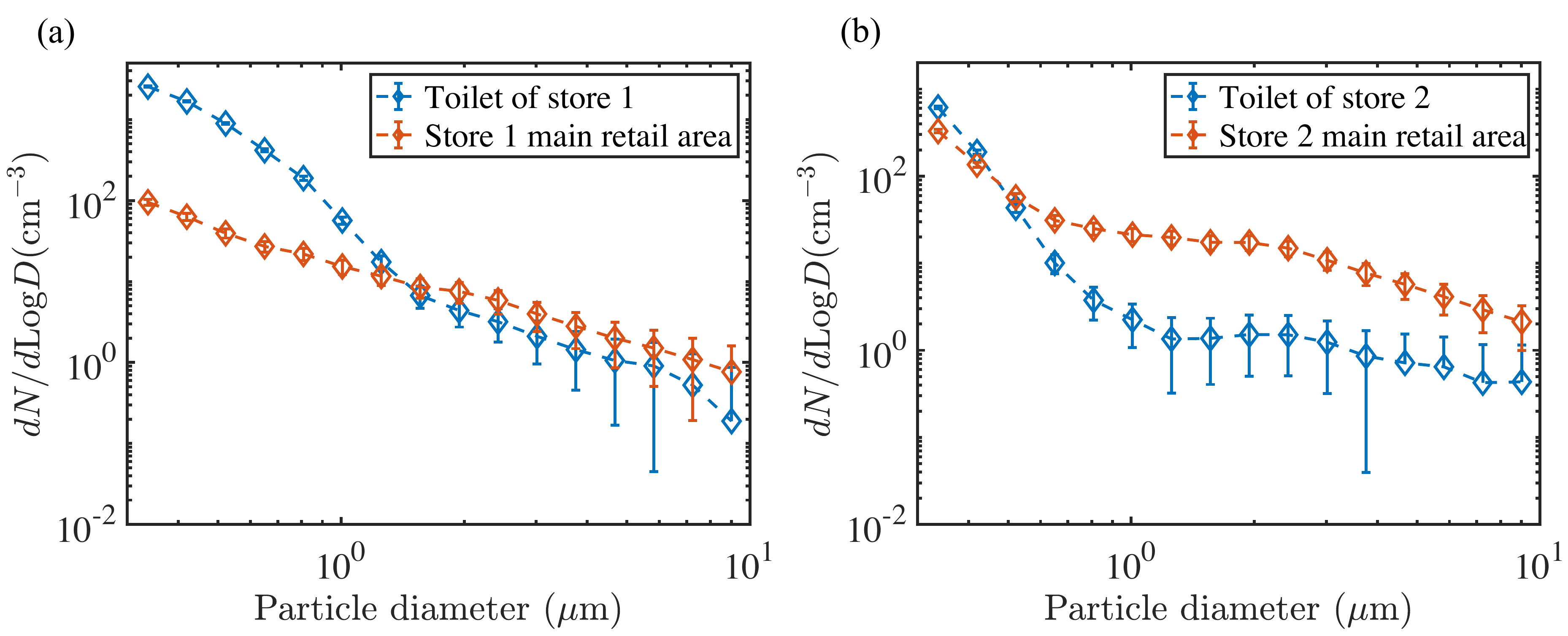}
	\caption{Background particle concentrations as a function of particle size for the toilets and main retail areas of (a) Store 1 and (b) Store 2.}
	\label{backgrounds}
\end{figure*}

The measured background particle concentrations for both stores as a function of particle size are shown in Fig.~\ref{backgrounds}. Let us first consider Store 1, for which the ventilation units circulate filtered outside air into the main retail area. As shown by Fig.~\ref{backgrounds}(a) for particle size $<\SI{1}{\micro\meter}$, the toilet had almost a factor 10 higher particle concentration than the retail spaces,  while for larger particles it was approximately the same. When we took the measurements, there was very little customer traffic in the toilet and thus little air exchange with the outside world through the entrance of the toilet. The strong increase for particles $<\SI{1}{\micro\meter}$ shows an accumulation of small aerosols and zero to little air exchange. The latter was confirmed by the lack of observed flow at the inlets/outlets at the ceiling.

Store 2 was different not only in layout of ventilation units, but also in that the fresh air in the main retail areas was unfiltered and thus the particle concentration in the retail areas was higher than in Store 1. The lower concentration in the customer toilet of Store 2 compared to the retail spaces of Store 2 for particles $>\SI{0.5}{\micro\meter}$ suggests air exchange with filtered air in the toilet. 

\textbf{Mean aerosol decay times}

The mean measured decay times, $\Bar{\tau}$ and mean 90\% aerosol dwell times, $\Bar{t}_{90\%}$ are summarized  in Table~\ref{table:exp}. The location for each experiment in the store is identified with the label S for store number and L for location number as shown in Fig.~\ref{locations}. 
$\Bar{\tau}$ and $\Bar{t}_{90\%}$ give the arithmetic mean of decay and dwell times measured over all bins of all spectrometers. The range for $\tau$ is given by $\frac{\tau_{min}}{\Bar{\tau}}$ and $\frac{\tau_{max}}{\Bar{\tau}}$ by  dividing the minimum and maximum  decay time in each experiment for different bins of the spectrometers by $\Bar{\tau}$. The mean  aerosol advection velocity $\Bar{v}$ for all spectrometers in an experiment for the 0.90-\SI{1.12}{\micro\meter} sized particles are also given in Table~\ref{table:exp}. The 0.90-\SI{1.12}{\micro\meter} sized particles were chosen as the data has the best signal-to-noise ratio.

\begin{table*}[!htb]
\fontsize{10}{6}\selectfont
\centering

\begin{tabular}{c c c c c c c}
 \multicolumn{6}{c}{} \\
\hline
\\
Exp. no. & Loc. label&
$\Bar{\tau}$ (min) &
$\frac{\tau_{min}}{\Bar{\tau}}$ and $\frac{\tau_{max}}{\Bar{\tau}}$ &
$\Bar{t}_{90\%}$ (min) &
$\Bar{v}$ (cm/s) &
Description
\\
\\\hline

\\
 
\textbf{1} & \textbf{S1 L1} & \textbf{34.37 $\pm$ 4.94} & \textbf{0.40-2.51} & \textbf{32.68 $\pm$ 0.65} & NA & \textbf{Toilet}
\\
\\\hline
\\
2 & S1 L2 &  1.13 $\pm$ 0.05 & 0.72-1.76 &3.35 $\pm$ 0.1 & 4.7 & Area of interest \\
3 & S1 L2 &  1.47 $\pm$ 0.03 & 0.87-1.44 & 4.13 $\pm$ 0.16 & 5.9 & \ditto \\
4 & S1 L2 &  5.82 $\pm$ 0.31 & 0.57-2.58 & 10.58 $\pm$ 0.1 & 3.3 & \ditto \\
5 & S1 L2 & NA & NA & 2.01 $\pm$ 0.03 & NA & Area of interest, brief event \\\\\hline\\
6 & S1 L3 & NA & NA & 3.07 $\pm$ 0.01 & 9.7 & Out of area of interest \\\\\hline\\
7 & S1 L4  & 1.49 $\pm$ 0.12 & 0.26-3.26 & 2.29 $\pm$ 0.09 & 24.3 & Under inlet \\
8 & S1 L4  & 1.26 $\pm$ 0.11 & 0.42-3.07 & 2.6 $\pm$ 0.11 & 12.7 & Under inlet \\\\\hline\\
9 & S1 L5  & 3.78 $\pm$ 0.14 & 0.86-1.39 & 6.6 $\pm$ 0.15 & 6.3 & Between two inlets \\
10 & S1 L5 & 2.49 $\pm$ 0.07 & 0.72-1.70 & 5.37 $\pm$ 0.05 & 2.8 & Between two inlets \\\\\hline\\
11 & S1 L6 & NA & NA & 2.37 $\pm$ 0.06 & 4.4 & Main path \\\\\hline\\
12 & S1 L7 & NA & NA & 7.05 $\pm$ 0.16 & 3.6 & Under outlet \\
13 & S1 L7 & 2.10 $\pm$ 0.06 & 0.85- 1.30 & 5.3 $\pm$ 0.08 & 2.2 & Under outlet \\\\\hline\\
14 & S1 L8 & 0.55 $\pm$ 0.04 & 0.81-1.85 & 2.16 $\pm$ 0.05 & 4.9 & Cash register \\
15 & S1 L8 & NA & NA & 2.08 $\pm$ 0.02 & 5.7  &\ditto \\
16 & S1 L8 & 0.66 $\pm$ 0.02 & 0.89- 1.21 & 2.09 $\pm$ 0.04 & 11.5 & \ditto \\\\\hline\\
17 & S1 L9 & 1.21 $\pm$ 0.04 & 0.80-1.46 & 1.84 $\pm$ 0.07 & 3.7 & Garden center \\\\\hline\\
18 & S1 L10 & 1.37 $\pm$ 0.10 & 0.51-1.72 & 4.5 $\pm$ 0.07 & 3.8 & Aisle to aisle \\
19 & S1 L10 & 1.55 $\pm$ 0.02 & 0.96-1.14 & 6.14 $\pm$ 0.05 & 7.2 & Aisle to aisle \\\\\hline\\
avg. & - & 1.91 & - & 4.08 & 6.86 & Store 1 average\\\\\hline\\

\textbf{20} & \textbf{S2 L1} & \textbf{9.34 $\pm$ 0.93} & \textbf{0.75-2.36} & \textbf{14.59 $\pm$ 0.09} & NA & \textbf{Toilet} \\
\textbf{21} & \textbf{S2 L1} & \textbf{10.74 $\pm$ 1.79} & \textbf{0.71-3.37} & \textbf{11.68 $\pm$ 0.09} & NA & \textbf{Toilet} \\\\\hline\\

22 & S2 L2 & 2.40 $\pm$ 0.22 & 0.49-3.47 & 4.62 $\pm$ 0.07 & 7.7 & High-traffic area \\
23 & S2 L2 & 1.37 $\pm$ 0.13 & 0.51-4.56 & 3.61 $\pm$ 0.09 & 8.2 & \ditto \\
24 & S2 L2 & 1.48 $\pm$ 0.07 & 0.85-2.06 & 3.69 $\pm$ 0.06 & 4.5 & \ditto \\
25 & S2 L2 & 2.46 $\pm$ 0.26 & 0.71-3.61 & 5.7 $\pm$ 0.19 & 5.9 & Aisle to aisle \\
26 & S2 L2 & 2.67 $\pm$ 0.22 & 0.73-2.1 & 5.97 $\pm$ 0.07 & 5.2 & Across main aisle \\\\\hline\\
27 & S2 L3 & 0.83 $\pm$ 0.07 & 0.66-4.32 & 1.65 $\pm$ 0.07 &  34.9 & Downstream of inlet\\
28 & S2 L3 & 1.12 $\pm$ 0.11 & 0.79-2.5 & 2.4 $\pm$ 0.07 & NA & Downstream of inlet\\
29 & S2 L3 & 0.58 $\pm$ 0.04 & 0.82-1.84 & 5.39 $\pm$ 0.03 & 25.7 & Downstream of inlet \\\\\hline\\
30 & S2 L4 & NA & NA & 1.72 $\pm$ 0.06 & 10.4 & Cash register, brief event \\
31 & S2 L4 & 1.96 $\pm$ 0.19 & 0.44-2.55 & 3.1 $\pm$ 0.04 & 7.7 & Cash register \\\\\hline\\
32 & S2 L5 & NA & NA & 0.5 $\pm$ 0.13 & 10.7 & Garden center, brief event \\
33 & S2 L5 & 0.59 $\pm$ 0.09 & 0.76-3.05 & 1.52 $\pm$ 0.06 & 16.5 & Garden center \\
34 & S2 L5 & 1.29 $\pm$ 0.14 & 0.66-2.79 & 5.11 $\pm$ 0.03 & 11.7 & Garden center \\\\\hline\\
avg. & - & 1.54 & - & 3.46 & 12.42 & Store 2 average \\

 & & & & &\\
 
\hline
\end{tabular} 
\caption{Mean aerosol decay times $\Bar{\tau}$, range of aerosol decay times $\frac{\tau_{min}}{\Bar{\tau}}$ and $\frac{\tau_{max}}{\Bar{\tau}}$ where we divide the minimum and maximum measured decay time in each experiment across all bins of the spectrometers by $\Bar{\tau}$, and mean 90\% aerosol dwell times $\Bar{t}_{90\%}$ for experiments performed in the two stores. The mean aerosol advection velocity $\Bar{v}$ for each experiment is also reported. Here S refers to the store and L to the locations identified in Fig.~\ref{locations}. The measurements made in the well-mixed environment of the customer toilets are shown in bold text. Store 1 and 2 averages include all locations except for the customer toilets and NA values.}
\label{table:exp}
\end{table*}

\textbf{Aerosol decay times with customer traffic}

In experiments 22-24, four to five volunteers were asked to move within the trial area during the measurement and behave in a similar way to the customers. For example, the subjects walked past the spectrometers, stopped at certain products to inspect them, and then moved on to another section of the aisle. We observed fluctuations in aerosol concentration at the spectrometer closest to the subject, while the other spectrometers showed little or no  fluctuations.  In the fluctuating areas smallest particles had  longer decay times, which is consistent with the expectation that they are more easily transported by the turbulent wake created by a walking person. Although the presence of the people created fluctuations that led to slight variations in the decay times, the decay times were very close to the store average at 1-2 minutes and the dwell times were very close to the store average at 3-5 minutes. We attribute this to the persistent airflow measured in the stores.  Additional information on the influence of people on the aerosol concentration measurements can be found in Fig. 9 of the Supplementary Material.

\textbf{Aerosol decay in the vicinity of ventilation inlets and outlets}

 Experiments 7 and 8 (see Supplementary Material, Fig. 2) were conducted directly under an \textit{air inlet} in \textbf{Store 1}, where the flow velocities at 1.5 m above the floor  directly at the inlet location were $\sim$ \SIrange{20}{50}{\centi\meter\per\second}, measured with a testo 405i thermal anemometer probe. As shown in Figure~\ref{locations}a, the ventilation units were mounted on the ceiling, with fans blowing filtered air downwards. Dust was released directly below an inlet. The dynamics of the aerosol decay times were strongly dependent on the distance of the spectrometers from the inlet area with $\tau_{min}<0.5\bar{\tau}$ and $\tau_{max}>3\bar{\tau}$. An extremely rapid drop in aerosol concentration was measured for the spectrometers nearest to the air inlet, which increased with distance from the air inlet. We attribute this increase above the store average to the recirculation of the room air driven by the rapid inflow below the air inlet.  This is even more evident in experiments 9 and 10, where the spectrometers were placed between two air inlets (see Supplementary Material, Fig.~3). The average decay times were even longer, which we attribute to the accumulation of aerosols in the stagnation region between the two recirculations driven by  the two air inlets. This is also evident from $\Bar{t}_{90\%}$, where the mean dwell times in experiments 7 and 8 are much shorter at 2.29 min and 2.6 min  than in experiments 9 and 10 at 6.6 min and 5.37 min between the two inlets. Experiments 7 and 8 had higher mean velocities of $\Bar{v}_{exp.7}=24.3$ cm/s and $\Bar{v}_{exp.8}=12.7$ cm/s than experiments 9 and 10 with $\Bar{v}_{exp.9}=6.3$ cm/s and $\Bar{v}_{exp.10}=2.8$ cm/s. The \textit{air outlets} did not significantly influence aerosol dynamics since the measured $\Bar\tau$ in Experiment 13 is 2.1 min and close to the store average of \SI{1.91}{\minute} (and $\Bar{t}_{90\%}=5.3$ min, with store average being 4.08 min). The air outlets also don't appear to create much air flow as $2$~cm/s~$<\Bar{v}<4$~cm/s and is less than the store average of \SI{6.86}{\centi\meter\per\second}.
Measurements near ventilation units in \textbf{Store 2} gave similar results.  There the ventilation units were installed on walls and fans blew unfiltered outside air into the store. The  experiments performed downstream of the inlet in Store 2, Experiments 27-29 (see Supplementary Material Fig.~10 for details on the experiment site), had relatively short decay times due to strong turbulent advection from the large scale flow ($0.58$~min~$<\Bar{\tau}<1.12$~min, which are less than the Store avg. of 1.54 min), even though the range was still relatively broad with Experiment 27 having a $\tau_{min}=0.66\Bar{\tau}$ and $\tau_{max}=4.32\Bar{\tau}$. Experiments 32-34 were also performed in front of air inlets in the garden center of Store 2 which show similar behavior to the main retail areas in terms of their decay times ($0.59$~min~$<\Bar{\tau}<1.29$~min). 
These locations also had short event duration in comparison to other locations in Store 2. The exceptions for these group of experiments are Experiments 29 and 34, where the spectrometers were intentionally moved to regions where the aerosol cloud was advected to by the mean air flow created by the inlets, and as a result, we measure longer $\Bar{t}_{90\%}$ since the aerosol cloud has diffused significantly and spread-out spatially by the time it reaches the spectrometer locations, so while the event duration is longer the particle concentration is lower due to dilution with the ambient air ($\Bar{t}_{90\%}$ in these locations is larger than the store average of 3.46 minutes).
The flow from the inlets also cause the mean advection velocity in these regions, $10.7$~cm/s~$<\Bar{v}<34.9$~cm/s, to be approximately the same or higher than the store average of $\Bar{v}=\SI{12.42}{\centi\meter\per\second}$.

\textbf{Aerosol decay in areas of long customer dwell times}

\begin{figure*}[htbp]
	\centering
	\includegraphics[width=0.9\textwidth]{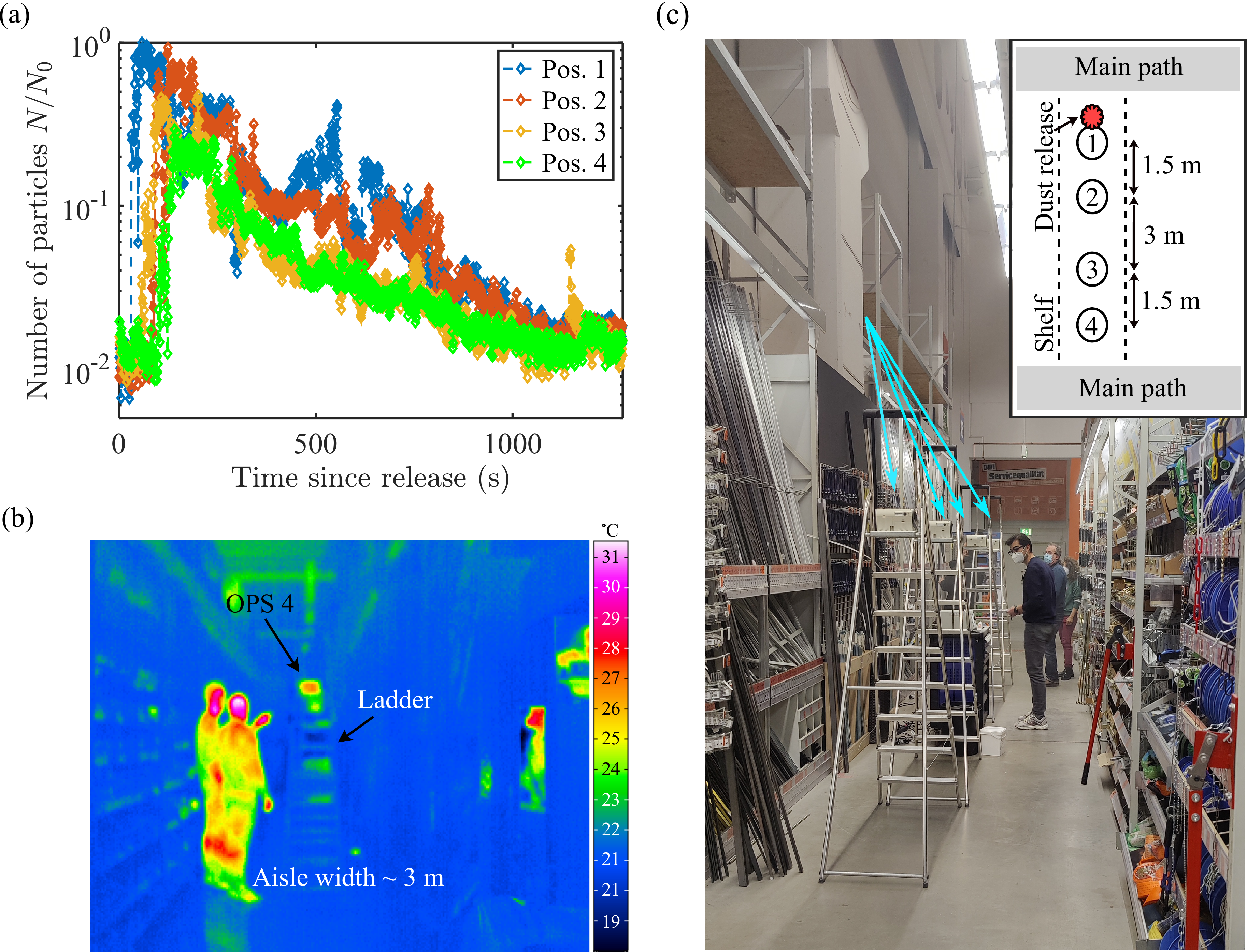}
	\caption{(a) Normalized aerosol concentration for particles of size 0.90-\SI{1.12}{\micro\meter} at different positions in the aisle for Experiment 4 (S1 L2). (b) Thermal camera image of an aisle with two people standing in the aisle. The temperatures in the store are relatively constant with no significant contributions to aerosol transport from thermal convection. (c) Schematic of the experiment site along with a photograph of the aisle with spectrometers placed on ladders and researchers present. Main path refers to areas with high movement where people move from aisle to aisle.}
	\label{screw thermal}
\end{figure*}

Experiments 2-5 were conducted in the nuts, bolts, and screws aisle in \textbf{Store 1}, an area of interest to store operatives as customers have a longer dwell time there. Experiments 2 and 3 show short decay and dwell times with $\Bar{\tau}$, $\Bar{t}_{90\%}$, and $\Bar{v}$ are close to the store averages. However in Experiment 4 we measured the longest decay and dwell times of the main retail areas of both stores. With $\Bar{t}_{90\%}=10$ min was more than twice of that  measured in the same location (S1 L2) in other experiments. In that case the  measured mean advection velocity was  $\Bar{v}=3.3$ cm/s, which is nearly a factor 2 less than the store average of $\Bar{v}=6.86$ cm/s. We attribute this slower velocity to changes in the large scale flow due to opening and closing of nearby entrance/exit doors. This factor of two in decay times was measured only once in this experiment and even then it is still quite fast with a 10 min decay. Details of Experiment 4 are shown in more detail in Fig.~\ref{screw thermal}.

The majority of experiments performed in the open store areas were areas where the aerosol dynamics was dominated by advection and diffusion due to turbulent mixing and sweeping flows.
In Fig.~\ref{screw thermal}(a), after the release of the aerosol, the short delay between rise in concentration levels between spatially separated spectrometers show the fast transport of aerosols caused by turbulent advection.
Furthermore, the aerosol concentration decays at the same time at all 4 positions meaning that turbulent mixing rapidly spreads the aerosol in the aisle and advection removes the aerosol from the environment.
Additionally, there is no significant contribution to aerosol transport from thermal convection as demonstrated by a thermal image of the experiment aisle shown in Fig.~\ref{screw thermal}(b). The temperature profile of the store is relatively constant with no gradients between the floor, ceiling, and walls that would cause convective motion of air and aerosols.
Figure~\ref{screw thermal}(c) shows the schematic of the experiment site along with a photograph of the aisle displaying the 4 spectrometers that are placed on ladders while researchers in the aisle monitor spectrometer readings.

\textbf{Aerosol decay times at the cash registers}

Interestingly, the cash register areas of both stores, which are close to the main entrance/exit doors of stores, also have relatively short $\Bar{\tau}$ and $\Bar{t}_{90\%}$ as compared to the store averages. 
For Store 1 these values are, $\Bar{\tau}\sim0.5$~min and $\Bar{t}_{90\%}\sim2$~min, and for Store 2, $\Bar{\tau}\sim2$~min and $\Bar{t}_{90\%}\sim2-3$~min.
These regions can be of importance in regards to the infection risk of customers and store personnel since they can have a higher density of people present at checkout lines as compared to other locations in the store. In our experiments however, the presence of people in the checkout area was minimal due to store capacity limits and since the measurements were done at closed cash registers.
While the measured characteristic times are relatively short, $\Bar{v}$ is not consistent between different experiments in the same locations for the cash register experiments and varies between 5 and 11 cm/s for Store 1 and 7 to 10 cm/s for Store 2. We contribute the difference in $\Bar{v}$ to the changing flows created by the opening and closing of the entrance/exit doors of the stores. At the same time, the opening and closing of doors and the air exchange with outside air assists in the rapid removal of aerosols and thus explain the short measured decay times and event duration.
The garden center of Store 1 also has short $\Bar{\tau}=1.21$ min and $\Bar{t}_{90\%}=1.84$ min, which is most likely due to the open layout of the greenhouse and close proximity to service doors used for moving inventory, as shown in Fig.~\ref{locations}. The mean advection velocity in this area was 3.7 cm/s.
We have also examined aisle to aisle transport of aerosols in Experiment 18 and 19. These experiments have values of $\Bar{\tau}\sim$1-2 minutes and close to the store average. We find that the aerosol transport to neighboring aisles is not significant and that the shelving acts as a barrier to transport. Further details of the aisle to aisle experiments can be found in Fig.~7 of the Supplementary Material.

\textbf{Aerosol concentration decay in open spaces is by turbulent advection}

\begin{figure}[htbp]
	\centering
	\includegraphics[width=0.5\textwidth]{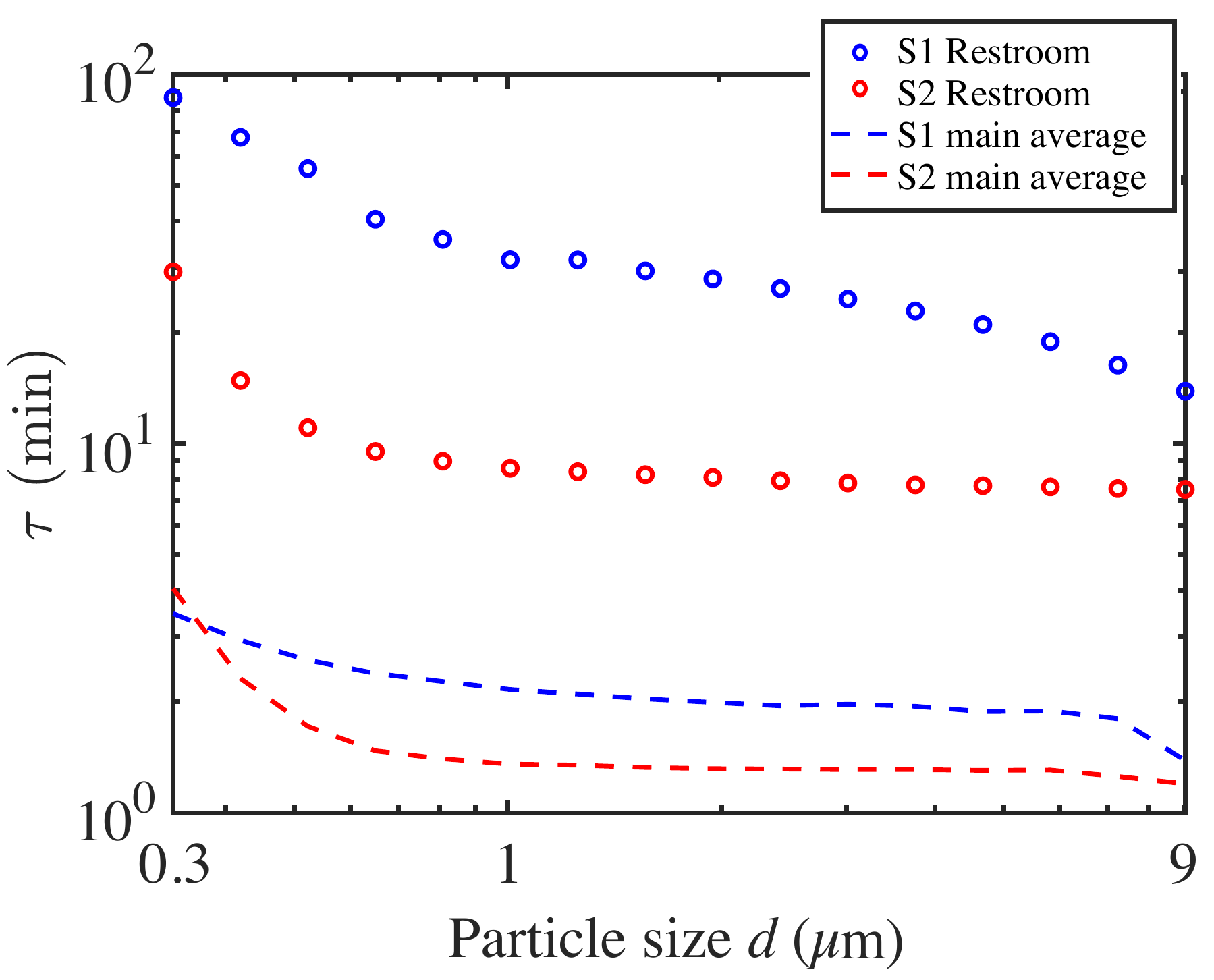}
	\caption{Aerosol decay times  $\tau$ as a function of particle size at different locations. In S2 main the ventilation intake air from outside was not filtered, leading to larger decay times for small aerosols.}
	\label{tau all}
\end{figure}

In order to better characterize the difference between open spaces and small well-mixed rooms we also measured the customer toilets of the two stores. Figure~\ref{tau all} shows the measured aerosol decay times for the different measurement locations as a function of particle size. Fig.~\ref{tau all} confirms the observed low air exchange of the toilet of Store 1 where the measured decay times are extremely long in comparison to other locations, commensurate with what one would expect for pure gravitational particle settling and deposition on the walls as shown by experiments in room sized chambers~\citep{Lai2004} (approximately $>\SI{10}{\minute}$ for most cases).
While in Store 2 the particle decay times in the customer toilets are commensurate with approximately 5-6 times ACH for particles $>\SI{0.5}{\micro\meter}$. This is in  agreement with German law that requires ventilation systems to have at least 5 times ACH for customer toilets~\citep{toilet_law}. The longer decay times for particles $<\SI{0.5}{\micro\meter}$ in the toilet of Store 2 could be attributed to the filter properties of the ventilation unit for the smaller particles.

In comparison, for the large open retail spaces the decay times are of the order of a few minutes and only weakly dependent on particle size. The uniformly short decay times for all measured aerosol sizes in the main open retail areas show that, unlike in the toilets that can be considered well-mixed, the aerosol removal mechanism in the open areas is not dominated by deposition and settling, which is a strongly size dependent process~\citep{Nazaroff2021}, but rather by transport through turbulent sweeping flows that remove contaminants.  The latter will only show a weak dependency on particle size as the decay is fast compared to the settling times of the particles studied.  

\begin{figure}[htbp]
	\centering
	\includegraphics[width=0.5\textwidth]{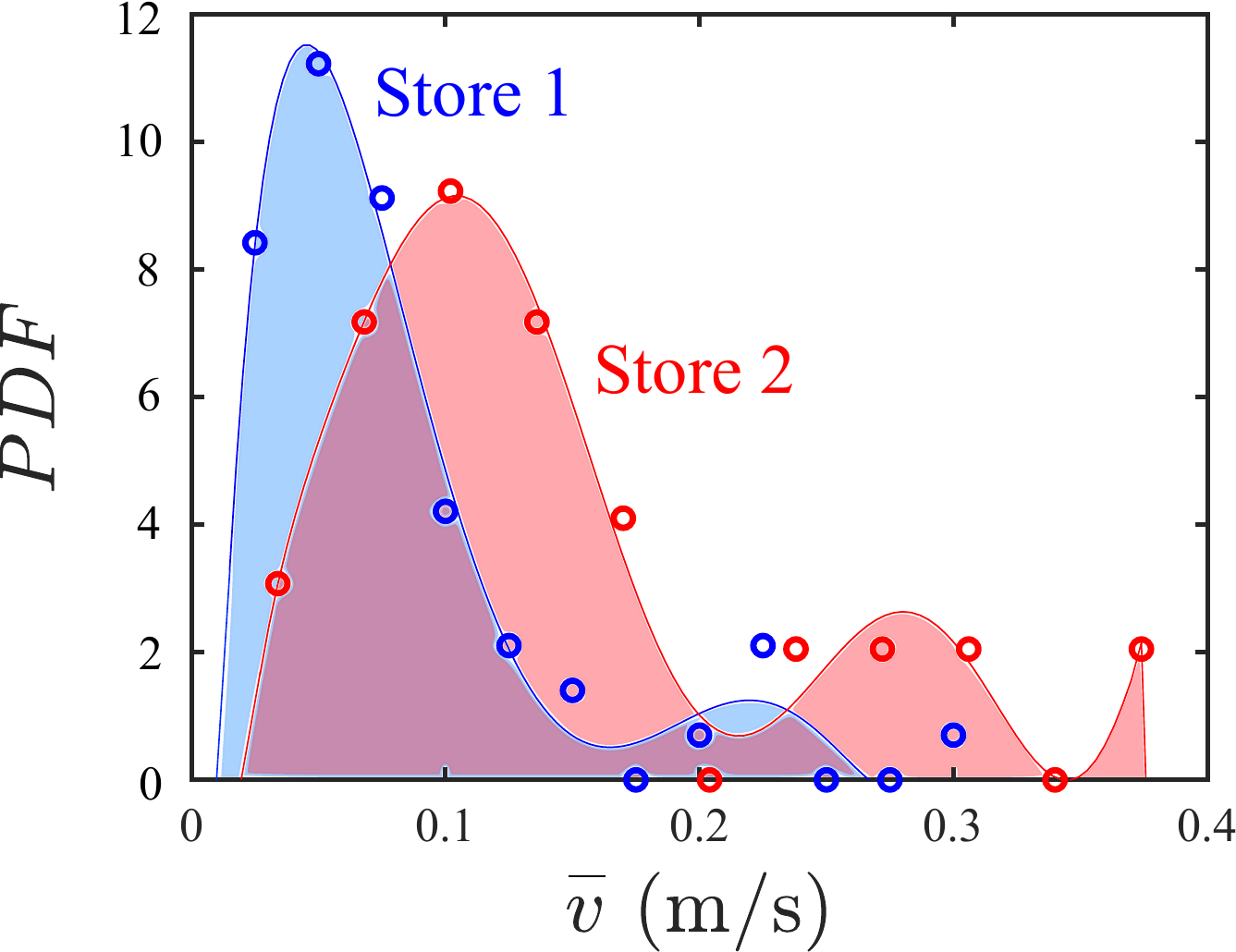}
	\caption{Probability distribution function (PDF) of the mean measured aerosol advection velocity in each experiment for Store 1 and 2. The shaded regions are polynomial fits to the data.}
	\label{v pdf}
\end{figure}

The longest average decay times were measured in the experiments with the smallest measured flow velocities, such as the experiments between the two inlets (Experiments 9 and 10), all other experiments while showing variations in the measured decay time have an average decay $\Bar{\tau}<$3 minutes with the average of the stores being between 1-2 minutes. 
Figure~\ref{v pdf} shows the probability distribution function (PDF) of the mean measured advection velocity $\Bar{v}$ in each experiment. We were able to calculate the aerosol advection velocity using the distance to the aerosol source and the 5\% arrival times for 57 spectrometer locations in Store 1 and 39 velocities in Store 2. The shaded regions are obtained by a polynomial fit to the data. 
The overall higher average velocity in Store 2 agrees with the average shorter decay times and dwell times and slightly higher nominal ACH of Store 2.
While the the aerosol is mostly advected with a velocity on the order of a few centimeters per second, which is typical of indoor spaces with air speeds $\sim$1-10 cm/s~\citep{Nazaroff2021}, however, in our experiments $\Bar{v}$ is persistent and directional as evident by experiments where the spectrometers are more than 10 meters away from the release location and yet the aerosol is detected by the spectrometers.
such as Experiments 27-29, and 32-34. 
The measured advection velocities are consistent with the transport of particles with a size of $<20$ $\mu$m being dominated by the room air flow rather than gravity, which is $<1$ cm/s for particles with a size of $<20$ $\mu$m~\citep{Nazaroff2021}.

\section{Conclusions}
In this study, we investigated aerosol transport in two German hardware/DiY stores with large retail spaces ($V>\SI{10000}{\cubic\meter}$).  The size distribution of the test aerosol used was similar to that of human exhaled aerosols. We focused on particle sizes in the range of $0.3$-$\SI{10}{\micro\meter}$, which are known to remain airborne for long periods of time. We conducted measurements in various locations that pose a high risk of infection for airborne diseases, such as frequently visited areas in large open spaces or small rooms such as customer toilets. We released the test aerosol locally and measured the decay and dwell times spatially resolved. Although the stores had different ventilation systems, in the large retail spaces the measured average decay times ($\bar{\tau}\sim$1-2 minutes) were short and only weakly dependent on particle size.  Even when we considered the total time that the aerosol remained at a given location, we found that the average measured aerosol dwell times of 4 minutes were only twice as long as the typical decay times. The fact that the decay and dwell times did not depend significantly on particle size suggests that turbulent transport by sweeping flows dominates over deposition. The measured advection velocities of the aerosols were very similar at all locations in the main retail areas of both stores, supporting this assumption.  The rapid decay due to sweeping flows also applied when people were in the measurement area. Although this led to temporal fluctuations, the measured average decay times did not change and were 1-2 min. These results need to be contrasted to the small ventilated spaces of the customer toilets, where the aerosol concentration decay times were $\bar{\tau}>10$ minutes. 

\textbf{Infection risk in the far-field is limited to the customer toilets}

The findings presented here have implications for disease transmission through infectious human aerosols. As mentioned, airborne transmission could occur either by far-field transmission, when the susceptible inhales infectious aerosols that have mixed and accumulated in indoor air, or by near-field transmission, when the susceptible is in the vicinity of infectious persons. Far-field transmission is limited to the customer toilets in the stores studied, where the risk of infection can be calculated by assuming a well-mixed room \citep{Nordsiek2021}. In the main retail areas, our results show that the accumulation of infection aerosols can be expected to be negligible. To better bound this, we consider an upper bound for the occupancy of such a store and the consequences for the accumulation of infectious aerosols. Given the typical exhalation rate of an adult person, which is \SI{0.5}{\meter^3} of air per hour when breathing, it should be safe when the exhaled aerosol concentration by all occupants is \textless 1\% of the volume capped at the height of a person ($\sim \SI{2}{\meter}$) for the contribution to aerosol accumulation in the room to be insignificant. This results in a maximum occupancy of \SI{25}{\square \meter}/person, which for the stores we studied means a maximum occupancy of 300-800 people in total (staff, customers, service staff, etc.) at any given time, which is well within the operating capacity of the stores we studied. 
In addition, it is important that there are no cluster infections among staff, which can be prevented for the example of the current COVID-19 pandemic by regular testing~\citep{Plank2021,Greenhalgh2022}. With these considerations, we can safely conclude that near-field one-to-one exposure is the most likely route of disease transmission in stores with large open retail spaces.

\textbf{Masks make the safe use of these large indoor spaces possible}

  In our recent work~\citep{Bagheri2021} we have shown that for the current variants of the SARS-CoV-2 virus, unprotected inhalation of infectious exhale has a more than 90\% risk of infection after only a few minutes, even at a distance of 3m.  Direct infection in the near-field is therefore extremely likely. Consequently, use of the stores would not be advisable without effective prevention measures, such as universal masking as shown in ~\citet{Bagheri2021}. With a well-fitting FFP2/KN95 mask (fitted manually without fit testing) worn by all parties, the conservative estimate of infection risk is $\sim$0.4\%, even after an hour of infectious speaking and in the worst-case scenario when the susceptible inhales the full undiluted concentration of aerosols exhaled by the infectious. As we pointed out in ~\citet{Bagheri2021}, this very conservative assumption represents an upper bound, and it is reasonable to assume that the actual risk can be a factor of 10 or even a factor of 100 lower, depending on the airflow at the mask. Let us now apply this to the situation considered here. If we assume a maximum 7-day incidence rate of about 2000 COVID-19 cases per 100000 households (1\% of the population with 2 persons per household), there are about 3 to 8 infectious persons in the store at any given time. (Please note that this value is intentionally set high).  An upper bound on the risk of infection can then be estimated by assuming that all the infectious people in the store are standing together and at the same time talking continuously to a susceptible in closest proximity for one hour. Even in this very extreme and unlikely situation, the upper bound of the risk of infection is only $\sim$3\% if everyone is wearing a well-fitting FFP2/KN95 mask. 
 
In conclusion, our data indicates that large indoor spaces similar to those studied are safe for employees and customers during the COVID-19 pandemic as long as (i) everyone wears a well-fitted FFP2 mask, (ii) the occupancy is not unbounded  and (iii) if there is no cluster infection between the employees.

\section*{Conflict of Interest Statement}
The authors declare that the research was conducted in the absence of any commercial or financial relationships that could be construed as a potential conflict of interest.

\section*{Author Contributions}

BH, OS, GB, EB: designing and carrying out the experiments; BH, OS, BT, GB, EB: analysis and interpretation of experimental data, writing original draft and final text; BH, OS, GB, EB contributed equally to this work.

\section*{Funding}
This work was partially funded by Handelsverband Heimwerken, Bauen und Garten (BHB).
Additional funding by BMBF within the project B-FAST (Bundesweites Netzwerk Angewandte Surveillance und Teststrategie) (01KX2021) within the NUM (Netzwerk Universitätsmedizin) and the Max-Planck-Gesellschaft. And also the Deutsche Forschungsgemeinschaft (project number: 469107130).

\section*{Acknowledgments}
The authors would like to thank Alexei Krekhov, Yong Wang, and Taraprasad Bhowmick for insightful discussions regarding solutions to the three-dimensional complex source advection-diffusion problem.

\section*{Supplemental Data}
 Details on all experiment sites showing aerosol release and spectrometer locations is shown in the Supplemental material along with further details on the aerosol release procedure. We have also provided a video of the aerosol release in the supplemental material.

\section*{Data Availability Statement}
The data presented here is available upon reasonable request.

\end{document}